\documentclass[english]{lni}
\usepackage{array}
\usepackage{multirow}
\usepackage{rotating}
\usepackage{comment}
\usepackage{lscape}
\usepackage{rotating}
\usepackage{longtable}
\usepackage{layouts}
\usepackage{graphicx}
\graphicspath{ {./images/} }
\usepackage{orcidlink}
\usepackage{amsmath} 
\begin{document}

\title[Potentials of Green Coding]{Potentials of Green Coding } 
\subtitle{Intermediate Report of the Project ''Potentials of Green Coding`` for the ISOC Foundation}

\author[Dennis M. Junger 
\and Max Westing 
\and Christopher P. Freitag 
\and Achim Guldner 
\and Konstantin Mittelbach 
\and Kira Obergöker
\and Sebastian Weber
\and Stefan Naumann
\and Volker Wohlgemuth]
{Dennis M. Junger \orcidlink{0000-0002-9271-2261}\footnote{HTW Berlin, University of Applied Sciences, Industrial Environmental Informatics Unit, Wilhelminenhofstr.~75A, 12459 Berlin, Germany \email{{dennis.junger|christopher.freitag|volker.wohlgemuth}@htw-berlin.de, Konstantin.Mittelbach@Student.htw-berlin.de}} 
\and
Max Westing \orcidlink{0009-0001-5044-7578}\footnote{Institute for Software Systems, Environmental Campus Birkenfeld, Trier University of Applied Sciences,55765 Birkenfeld, Germany
\email{{m.westing|a.guldner|s.naumann}@umwelt-campus.de}}
\and
Christopher Pascal Freitag \orcidlink{0009-0002-1611-024X}\footnotemark[1]
\and
Achim Guldner \orcidlink{0000-0002-7532-4523}\footnotemark[2]
\and
Konstantin Mittelbach \orcidlink{0000-0001-7590-2584}\footnotemark[1] 
\and
Kira Obergöker\footnotemark[2]
\and
Sebastian Weber \orcidlink{0000-0001-8260-2248}\footnotemark[2]
\and
Stefan Naumann \orcidlink{0009-0000-6542-2229}\footnotemark[2]
\and
Volker Wohlgemuth \footnotemark[1] 
}

\startpage{1} 
\editor{Gesellschaft für Informatik (GI)} 
\booktitle{ISOC - Potentials of Green Coding} 
\yearofpublication{2024}
\maketitle

\begin{abstract}

This document is the extended version of the literature report published at the German Informatik Conference titled "Potentials of Green Coding - Findings and Recommendations for Industry, Education and Science" \cite{junger23_litrep}. This document has since been updated and expanded to include relevant sources.
If you have any questions, input, or further ideas on the subject of green coding, do not hesitate to contact us at the indicated addresses.

Progressing digitalization and increasing demand and use of software cause rises in energy-
and resource consumption from information and communication technologies (ICT). This raises the
issue of sustainability in ICT, which increasingly includes the sustainability of the software products
themselves and the art of creating sustainable software. To this end, we conducted an analysis to gather
and present existing literature on three research questions relating to the production of ecologically
sustainable software (’Green Coding’) and to provide orientation for stakeholders approaching the
subject. We compile the approaches to Green Coding and Green Software Engineering (GSE) that
have been published since 2010. Furthermore, we considered ways to integrate the findings into
existing industrial processes and higher education curricula to influence future development in an
environmentally friendly way.

\end{abstract}

\begin{keywords}
Green Coding \and Potentials of Green Coding \and Curricula \and Software Engineering \and Literature Report \and ISOC Foundation \and German Informatics Society (GI) \and Trier University of Applied Sciences, Umwelt-Campus Birkenfeld \and HTW Berlin, University of Applied Sciences \and  Literature Analysis \and Workshop Review \and Project Review \and Pilot-Project \and State of the Art \and Green IT \and Green Computing \and Sustainable Software Engineering
\end{keywords}

\begin{figure}[ht]
    \centering
    \includegraphics[width=8cm]{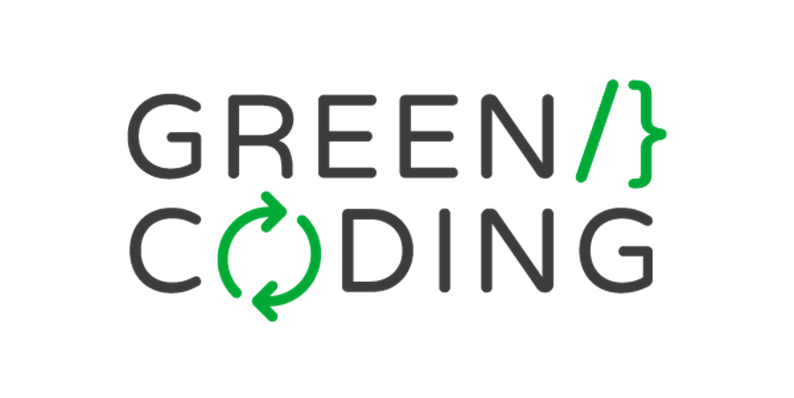}
    \caption{Logo of 'Potentials of Green Coding' project
(\url{https://gi.de/en/aktuelles/projekte/en-green-coding})}
    \label{fig:logo}
\end{figure}

\section{Introduction and Motivation}

Sustainability, generally relating to information and communication technology (ICT) and software development/engineering, is a research field of growing interest. While the focus of ecological sustainability considerations in ICT was on hardware for the longest time (Green IT), software behavior is the driver of hardware energy consumption and affects the hardware life-cycle, e.~g., as increasingly demanding software systems quickly outgrow older hardware forcing their replacement which consumes additional resources. To enable sustainable software production, access to models, methods, and tools geared toward sustainability is elementary. Furthermore, it is essential to make future IT professionals and researchers aware of the sustainability aspects of ICT and equip them with the skills to consider them in their work adequately. An area within this field is what we call 'Green Coding' (see definition below). The following report presents available Green Coding practices and information on how educators can incorporate them into their teaching.
\begin{figure}[ht]
    \centering
    \includegraphics[width=8cm, height=8cm]{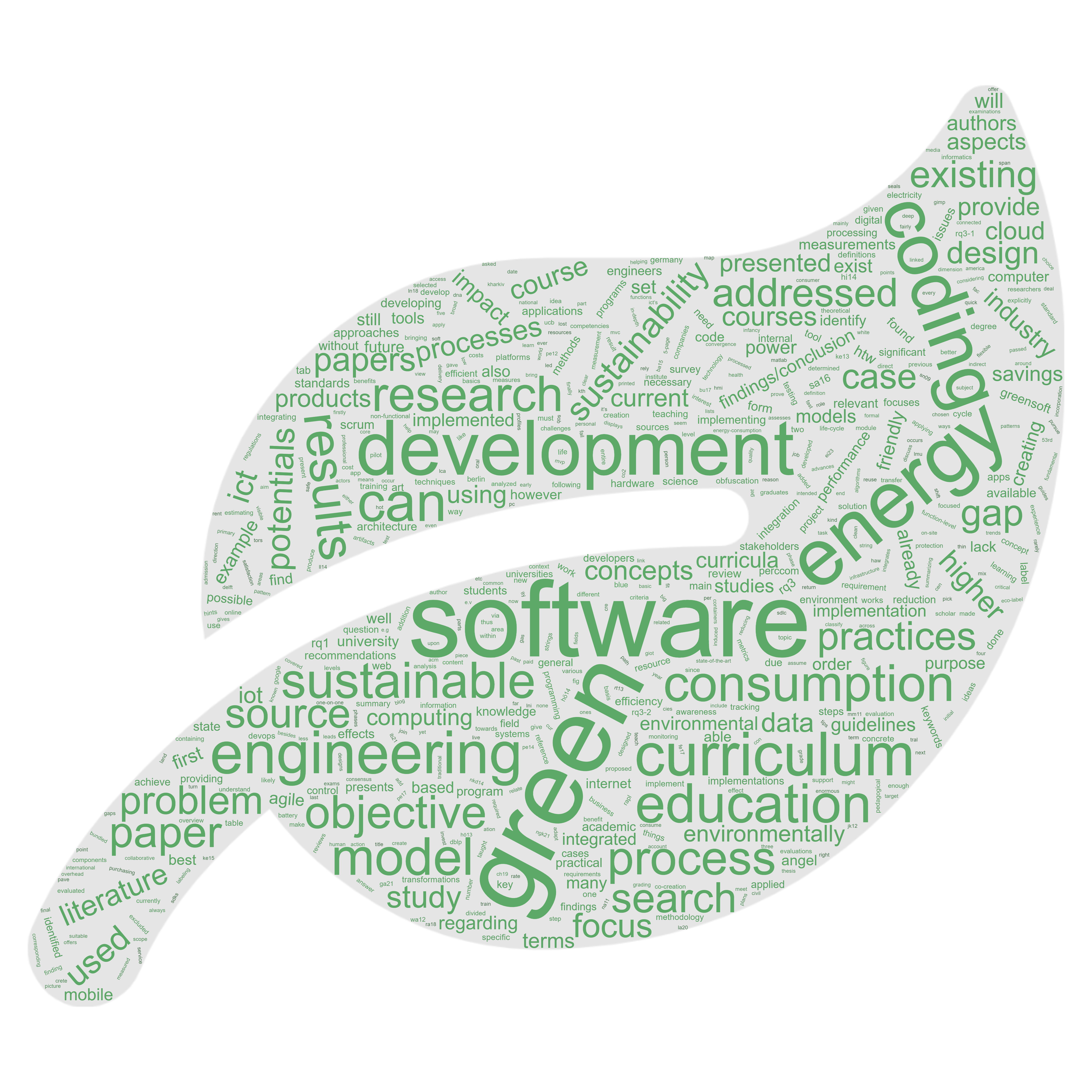}
    \caption{Weighted word cloud, created with \url{www.wordclouds.com}}
    \label{fig:wordcloud_label}
\end{figure}


\section{Used Definitions}
\subsection{Definition of Sustainability}

This report focuses on the environmental aspect of sustainability, dispensing with the other dimensions or pillars commonly defined as part of sustainability~\cite{Purvis_2018}. Other dimensions are, of course, also important and have been detailed e.~g. in \cite{Albuquerque2021} (social and technical), \cite{Johann2013} (social), \cite{Penzenstadler2013,Calero2017} (overview).

\subsection{Definition of Green Coding}

As it relates to ICT, research on sustainability is done under several terms, varied in definition and scope (e.~g. Green IT/ICT, Green by IT/ICT, Sustainable Software, etc.). The terms 'Green IT' and 'Green By IT' separate the field into two areas, 'Green By IT' designates means to implement solutions to increase the sustainability of other domains, using ICTs (e.~g. smart grids, building automation, environmental management information system (EMIS) like life-cycle-assessment software (LCA), etc.), whereas 'Green IT' deals with making ICT itself more sustainable (e.~g., follow-the-sun load shifting, backward compatible software, building the hardware in a way that it requires less energy and resources, etc.). 'Green Coding' is a part of 'Green IT' and focuses on all aspects of reducing the negative environmental impacts of ICT, which are triggered by the software product. For the purpose of this work, we define it as follows.
\begin{quote}
Green Coding is the act of designing, developing, maintaining, and (re-)using software systems in a way that requires as little energy and natural resources as possible. Green Coding methods or practices thus mean any action or use of technology intended and suitable to further this.
\end{quote}
Hence, Green Coding focuses on the methods and tools that can be leveraged during the production of a software product that impact the software induced ecological sustainability across all life-cycle phases (development, usage, and disposal).

\section{Research Questions}
To approach the topic, we conducted a literature analysis on three research questions relating to Green Coding.

\subsection{RQ1: What concepts of programming software in an environmentally friendly way exist for general software development?}

For this RQ, we gathered and present the most widespread Green Coding practices and tools from academic literature. Furthermore, we present and evaluate methods and tools, gathered in "best practice" lists. This should provide an overview to IT professionals who approach the topic of sustainable software development, aiding them to navigate the field and helping them to find Green Coding practices and tools applicable to their work.

\subsection{RQ2: What concepts of environmentally friendly software engineering processes are already in place in the Internet industry?} 
This question is intended to yield information on Green Coding practices used by software developers. Its target was to find case studies or white papers on implementing Green Coding practices and their impact on software performance and sustainability metrics.

\subsection{RQ3: How can Green Coding concepts be implemented into the current software development processes and the curricula of existing study programs?} 
Green Coding, primarily concerned with providing methods and tools to software developers and practitioners, plays a central role here. RQ3 is supposed to explore how Green Coding practices can be communicated to future IT professionals to establish competence in using available tools and methods in their training and work concepts.
To reduce the complexity of RQ3 and match the papers more precisely to their purpose, we divided RQ3 into two sub-questions, from education to skill enhancement.

\subsubsection{RQ3-1: How can Green Coding concepts be implemented into the current software development processes?} 
The focus of RQ3-1 is to figure out the possibilities of implementing new methods and tools into established software development models, processes, and tools. 

\subsubsection{RQ3-2: How can Green Coding concepts be implemented into the curricula of existing study programs?} 
To achieve the overarching goal of lowering the environmental impact of software, it is crucial to give future IT professionals the skills needed to produce sustainable software. Thus, the focus of RQ3-2 is to find possibilities for implementing the approaches into the education of future IT professionals. As most software developers have an academic background, this review focuses on implementing aspects of Green IT into the curricula of higher education institutions. The basis for this idea is a survey from 2022~\cite{statista22}, which describes the levels of formal education for software developers. In the survey, 41.32~\% of the participants stated to have a bachelor's degree, 21.14~\% have a master's degree, and 12.73~\% at least attended universities. Thus, 75.19~\% of the software developers have a higher education background.

\section{Search Design and Review Process}

The review was conducted as an exploratory search based on existing knowledge of papers as well as examinations on google scholar, dblp, and the online catalogs of the university libraries of the University of Applied Sciences Trier and the (HTW Berlin – University of Applied Sciences) for the terms 'sustainable software development processes', 'sustainability as non-functional requirement', 'creation of curriculum'. Papers using sustainability in a different context than defined above (i.~e., not focusing on the environmental dimension of sustainability) were excluded, as were papers from non-computer science fields (e.~g., results about green (building) code). This was decided based on the title, research area, and abstract. Some literature that would likely have been applicable had to be excluded based on access restrictions.

\subsection{Used Search Strings}
The following Tab.~\ref{tab:Seach_Strings} contains an overview of the search strings, the intended purpose behind the searches, as well as the amounts of search results.
\begin{longtable}[H!]{|m {3.55cm} |m {3.55cm}|m {1.1cm}|m {1.1cm}|m {1.1cm}|}
  \caption{Used keywords in review searches for queried databases}
  \label{tab:Seach_Strings}
  \small\\
  \hline
  \textbf{\raggedright Search String} & 
  \textbf{\raggedright Purpose} & 
  \textbf{\raggedright Google Scholar Results} & \textbf{\raggedright ACM DL Results} & 
  \textbf{\raggedright dblp \break Results} \\
  \hline 
Implementing green coding in software development &
To find information and guidance on integrating environmentally responsible practices into software development processes.& 
about \break  17800 &  
372203 & 
0 \\ \hline 

How to integrate sustainability in software development&
To learn the steps and best practices for making software development more sustainable.&
about \break 14900 &
384522 & 
0   \\ \hline

Steps to a greener software development process& 
To identify the steps that already exist for the integration of Green Coding into existing software development processes                               & about \break 17800  & 
384522 & 
0 \\ \hline

Best practices for implementing green coding in software engineering&
To identify the steps that already exist for the integration of Green Coding into existing software engineering processes&
about \break 16300 & 
378197 & 
0  \\ \hline

Strategies for implementing sustainable software development&
To identify the actual strategies for sustainable software development. &
about \break 17800 & 
373671 & 
0 \\ \hline

(green coding | green computing | green software engineering | sustainable software engineering) AND (practices | recommendations | techniques) &
Searching for concrete measures and guidelines  &
about \break 47400 & 
315 & 
0  \\  \hline

(green coding | green computing | green software engineering | Sustainable Software Engineering) AND (practices | challenges | review) &
Searching for existing reviews, summary sources, and research gaps &
about \break 20100 & 
314 & 
0  \\  \hline

(green coding | green computing | green software engineering | Sustainable Software Engineering) AND (case stud* | implement*) & Searching for implementations of or case studies on the usage of green coding practices &
about  \break 17800 &
317 &
0\\  \hline

(green coding | green computing | green software engineering | sustainable software engineering) &
Search string for dblp, as it did not return any results for the search strings including conjunction above &
about \break21900 &
317 &
72\\ 
\hline
\end{longtable}
\normalsize

Despite the many results delivered by the selected platforms since 2010, the authors noticed that only relatively few papers and articles found led to the desired results. In some cases, only 2-3 papers are relevant. Many of the results are not directly related to 'Green Coding' as a subcategory of 'Green IT' but have to do with non-relevant general topics due to the current prevalence of the words 'sustainable', 'green', 'coding', 'software', 'engineering' and 'development'. Because of this, it is not meaningful to use these numbers as statistics. The authors have thus agreed to pick out relevant sources individually and bring them together after discussing their relevance to get an overview of the field to be investigated meaningfully.

The many results are, as was observed, probably because the keywords such as 'Green', 'Coding', and 'Sustainability', as well as related ones, are nowadays in almost every paper and article and so significance in them is lost. As shown in the ACM DL Result column of \ref{tab:Seach_Strings}, the number of results varied enormously from using multiple-word phrases containing the desired goal to using Boolean operators for the combination of the selected keywords assumed to occur in relevant papers. As a result of this observation, the authors focused on the used Boolean combinations.

\begin{figure}
    \centering
    \includegraphics[width=\textwidth]{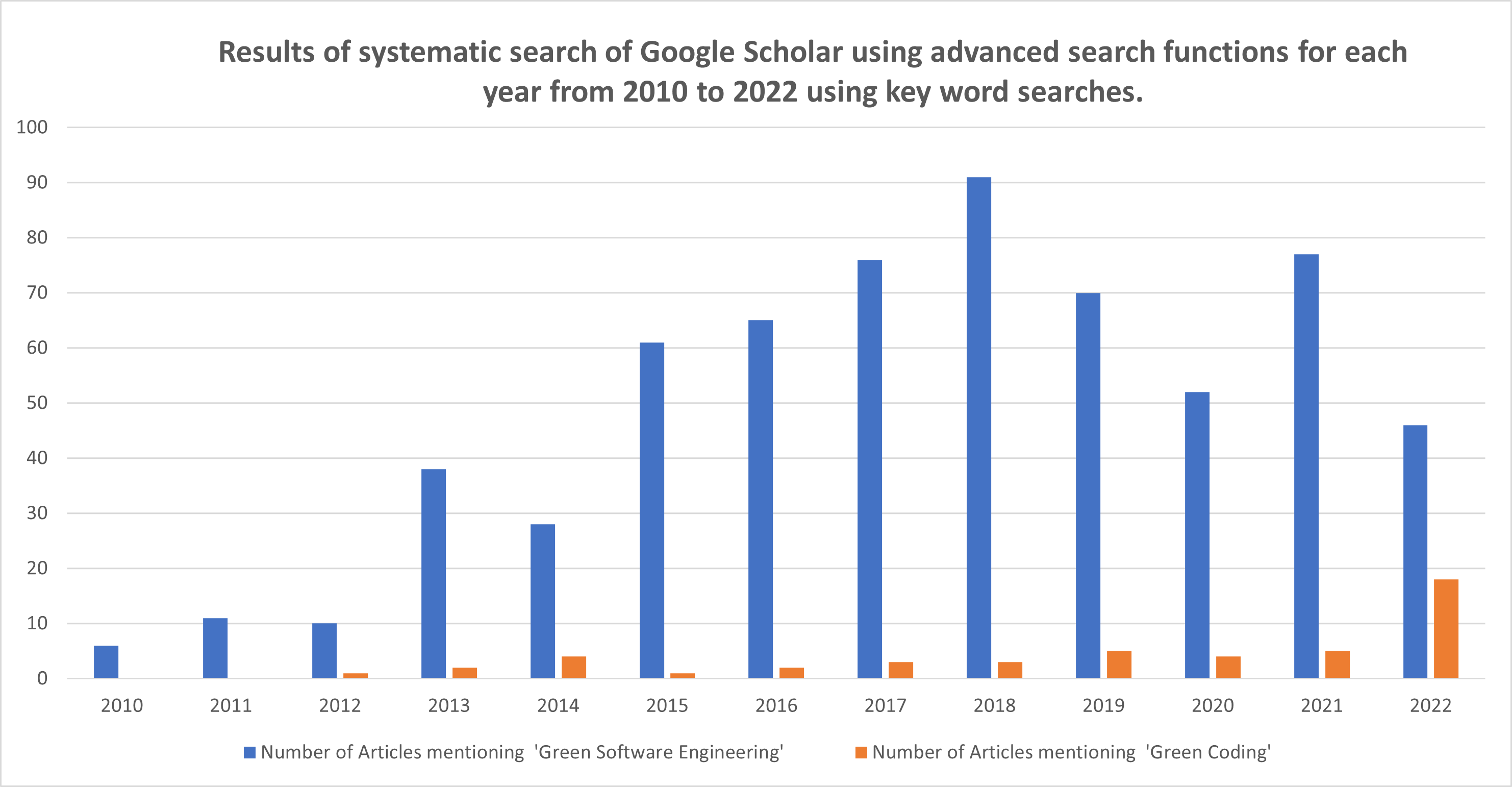}
    \caption{Results of a systematic search of Google Scholar using advanced search functions for each year from 2010 to 2022 using keyword searches}
    \label{fig:findings_per_year}
\end{figure}

In addition to using keywords such as 'Green Software Engineering', as explained in the introduction, 'Green Coding' is a primary keyword for the practices investigated in the project. Without technical expertise, it is difficult to identify relevant sources. However, combining the exact terms 'Green Coding' is also used in other scientific disciplines such as DNA sequencing and yields results from e.~g. psychology and economics and business administration on occasion. To collect a more meaningful data set for the following data displayed in Fig.~\ref{fig:findings_per_year} of the articles and papers per year, articles that do not fit the intended study objective were excluded.

\subsubsection{RQ1: What concepts of programming software in an environmentally friendly way exist for software development in general?}
The search strings for this RQ were supposed to produce literature on summary sources (i.~e., previous literature reviews or introductory texts), concrete measures and guidelines, and literature on implementations of Green Coding practices.

\subsubsection{RQ2: What concepts of environmentally friendly software engineering processes are already in place in the Internet industry?}
The search terms for RQ2 are not very different from those used in RQ1. In the choice of articles, attention was paid to the Internet reference. Explicit awareness was paid to IoT, cloud computing, and web references.

\subsubsection{RQ3: How can Green Coding concepts be implemented into the current software development processes and the curricula of existing study programs?}
The search terms used for this RQ are again similar to those used in the search terms of RQ1 and also in the printed Tab.~\ref{tab:Seach_Strings}, but in addition to the search for current techniques, the focus was placed on the practical usage already implemented.  

For the sub-question 'How can Green Coding concepts be implemented into the current software development processes?', in addition to the state-of-the-art techniques, additional techniques were searched. For this purpose, additional keywords like 'case study' were used, as also 'incorporation sustainability' or 'development models' were essential terms to relate the search results to the integration into the software development process. To reflect the different phases of software development and the necessity of sustainability in the software development process, we added the specific search for 'sustainability as a non-functional requirement'. Currently, we regard including sustainability as a non-functional requirement in the development process as the best formal integration of the topic into this early phase.

Compared to question 3.1, for the sub-question 'How can Green Coding concepts be implemented into the curricula of existing study programs?', the focus was set on the already implemented processes and the degree of freedom extending existing curricula in completely new ways. Accordingly, ICT curricula that do not currently represent any Green Coding approaches but are also integrated into the general curricula were selected. Thus, we appended the sustainability keywords with further search terms such as 'curriculum', 'higher education', 'creation of curricula', and 'curricula development'.

\pagebreak
\section{Processed Sources and findings}
\subsection{RQ1: What concepts of programming software in an environmentally friendly way exist for software development in general?}
Energy consumption at the hardware level is well-understood. However, the impact of software on energy consumption has only been regarded in the past decade and has gained more attention since about 2015. There are models for estimating or measuring the energy consumption of software and tools based on these models that seem suitable to inform software developers about energy optimization potentials in their code \cite{Kansal2010, kern13, hoenig13, hindle14, LeBeane2015, Fieni2020, Henderson2020, Aslanpour2021, Anwar2021, Schmidt2021, Khandelwal2022}. Furthermore, hardware-based power measurement tools like RAPL\footnote{\url{https://01.org/blogs/2014/running-average-power-limit-\%E2\%80\%93-rapl} and \url{https://cdrdv2.intel.com/v1/dl/getContent/671200} [2023-04-28]} or nvidia-smi\footnote{\url{https://developer.nvidia.com/nvidia-system-management-interface} [2023-04-28]} exist, that can be used to quickly assess the energy consumption of individual hardware components.

A common theme in many studies is that there rarely is a definitive solution that is always optimal concerning energy- and resource consumption \cite{peireira17}. Some studies like \cite{Jagroep16} compare hardware-based measurements with simultaneously obtained estimations and find that there are sometimes large (over 60\% in this case) differences in the measurements vs. the estimations due to unexplained energy overheads. This demonstrates the complexity of software sustainability and highlights why IT professionals need to be able to assess which measures are suited for a specific project. However, the steps required to improve upon a given piece of source code's induced energy consumption are not as clear. Some work provides concrete hints \cite{hoenig14} or details where the energy hot spots occur in the source code \cite{wang12, Noureddine2015, Verdecchia2018}.

There is research approaching software sustainability at various points of a product's life, at the design phase using software architecture (e.~g. \cite{chowdhury19}), during development (e.~g., \cite{hoenig14,wang12}) or maintenance (e.~g. \cite{sahin16,bunse17,calero21b}). Most of these approaches rely on energy measurements and a broad conceptual basis, such as the works of \cite{hindle14,kern13,naumann14}. Several studies focus explicitly on applications for mobile devices (e.~g., \cite{sahin16,bangash22,hindle14}), as energy consumption is of particular interest here due to its impact on battery life. 

Advances have been made in the development and application of standards to sustainable software products and labeling them \cite{kern18,naumann21}. However, no sustainability standards applying explicitly to software development have been identified by standards organizations. 

Guidelines or studies spanning the entire software development process and defining artifacts and tools exist, but usually operate on a higher level. They can e.~g. include models, process artifacts or evaluation methods, but not Green Coding methods such as coding guidelines or using specific architectures and technologies.

\begin{landscape}
\setlength\LTcapwidth{\linewidth}
\begin{longtable}[h]{|m {4em}|m {12em} |m {15em} |m {16em}|}
    \caption{Analyzed Papers regarding RQ1: What concepts of programming software in an environmentally friendly way\\(“Green Coding”) do exist for software development in general?}
    \label{tab:rq1_label}\\
    \hline
    \textbf{Source} & \textbf{Research Objective} & \textbf{Problem or Gap Addressed} & \textbf{Findings/Conclusion}\\
    \endfirsthead
    \hline 
    \textbf{Source} & \textbf{Research Objective} & \textbf{Problem or Gap Addressed} & \textbf{Findings/Conclusion}\\
    \endhead
    \hline 
     \cite{penz12} & Literature review & Providing a summary on research activity in the field & Few results for original scope, requiring the inclusion to work not strictly directed towards sustainable software engineering\\
     \hline
     \cite{ease14} & Literature review & Building on \cite{penz12} & Identified active research areas and authors, summarizing reviewed literature according to 7 research questions\\ 
     \hline
    \cite{lago22} & Literature Review & No secondary studies on sustainability in software architecture available & Few studies on environmental sustainability within software architecture, only two implementation-oriented studies in the ecological dimension found\\
    \hline
    \cite{sahin16} & Investigation of the impact of code obfuscation on mobile app energy consumption & Code obfuscation is used extensively for mobile apps, but its impact on energy consumption and battery life is largely unknown & Code obfuscation can have both a positive and negative impact on battery life, with a negative impact being more likely. This impact is not likely meaningful for the consumer \\ \hline
    \cite{bunse17} & Reproduction of \cite{sahin16} using some other obfuscation tools and set of apps & Extending previous findings to other cases to enable meta-analysis & Broadly reproduced the findings in \cite{sahin16} \\
    \hline
    \cite{bangash22} & Investigating the impact of byte-code transformations on app energy efficiency & Byte-code transformations are a common tool for app developers, but their impact on energy efficiency is unknown & Reducing Energy consumption with byte-code transformation is possible, but finding the right transformations is time-consuming. Applied transformations can also increase energy consumption\\ \hline 
    \cite{hoenig14} & Development of a framework that informs developers about the energy-consumption of their code & Platform-diversity necessitates a cross-platform approach to estimating energy consumption during development & Proposed frameworks which can provide accurate enough energy-consumption estimates across multiple platforms \\ \hline
    \cite{hoenig13} & Development of a tool that automates energy measurements and gives hints on how to optimize code for energy efficiency & Not enough easy-to-use tools are available to enable sustainable software engineering & Development of a tool that provides concrete hints to optimize code for energy-consumption \\\hline
    \cite{Mancebo2021} & Investigating the Link between software maintainability and energy consumption & Energy measurements of software between releases are uncommon & Identified a number of maintainability metrics correlating with software-induced energy consumption \\\hline
    \cite{wang11} & Creating a performance monitoring counter-based power model and a tool for estimating the energy consumption of software at function-level & To optimize code or energy efficiency, it is necessary to understand which piece of code induces which energy consumption & The created model and SPAN are suitable for providing function-level energy consumption estimates based off source code. \\ \hline
    \cite{wang12} & Creating an automated function-level power profiler & Manually creating power profiles for individual functions is time-consuming & Research objective is achieved. The resulting tool can be used to inform about software-induced energy consumption at the function level.\\ \hline
    \cite{hindle14} & Creating an energy-consumption focused testbed for android mobile devices & Mining software repositories are not yet used for energy consumption & Successful creation of a testbed for automated energy measurements via standard usage scenarios and measurement hardware \\ \hline
    \cite{peireira17} & Study on the relation between energy consumption and other software performance metrics for different programming languages & Choice of language might have effects on energy consumption & No consistently more energy efficient programming language found. Faster languages are not always more energy efficient.\\ \hline
    \cite{feitosa17} & Investigation of state/strategy and template method patterns impact on energy consumption by comparing them with alternative implementations fulfilling the same task & Design patterns are a mainstay in software developers' toolkits but are connected with additional overhead that may induce additional energy and resource consumption & The chosen alternatives outperformed the design pattern implementations in most cases. However, the pattern implementations were equally or more energy efficient in some cases \\ \hline
    \cite{guldner17} & Establishing how to measure and verify software sustainability criteria for standard software & Lack of standards for assessing the sustainability of software & Established a method to assess software sustainability based off energy consumption measurements and hardware utilization metrics\\ \hline
    \cite{naumann14} & Developing a life-cycle based reference model encompassing both GreenIT and Green by IT that can be implemented into new and existing development processes to enable sustainable software & A reference model enables stakeholders to relate information and artifacts to specific aspects of software sustainability & Proposal of the GREENSOFT model, a life-cycle based reference model containing a life-cycle model of software sustainability, criteria and metrics, procedure models, and recommendations and tools\\ \hline
    \cite{kreten18} & Investigating the energy-consumption of web apps deployed in containers. How does scaling affect energy consumption? & Lack of research on energy consumption behavior of containers & Containerization can achieve significant energy savings while increasing performance when scaling applications. Deployment of apps in a container without scaling them increases energy consumption. \\ \hline
    \cite{kern18} & Developing criteria and an evaluation method for assessing the sustainability of software products for the purpose of creating a software sustainability label & Lack of standards and recommendations for sustainable software. Software sustainability is not visible to consumers/customers & Produced a set of criteria and how to assess a software product in regard to them. \\ \hline
    \cite{chowdhury19} & Examining the energy consumption impact of Model-View-Controller (MVC) vs. Model-View-Presenter (MVP) architecture with bundled processing in mobile apps & Lack of awareness for 'green' software architecture & Software architecture can have a significant impact on energy consumption. Specifically, the change from MVC to MVP with bundled processing can achieve significant energy savings\\ \hline
    \cite{venters17} & Summary on the state of the art and research at the intersection of software architecture and sustainability & Challenges in software architecture appearing over the life-time of software might be tackled using the concept of sustainability & Providing orientation and structuring the research on sustainability and software architecture\\ \hline 
     \cite{Khandelwal2022} & Making energy consumption and carbon emission caused by ICT usage more transparent & Lack of data on carbon emissions caused by ICT & Provision of application for real-time carbon emission calculations \\ \hline
     \cite{Aslanpour2021} & Enabling management of resource-limited edge-computing nodes & Providing accurate and affordable measurements of energy consumption & Identification of energy-consumption factors, proposal and experimental validation of energy-consumption measurement framework WattEdge \\ \hline
     \cite{Kansal2010} & Creating power models for virtualized data centers & Difficult power consumption attribution in virtualized environments over traditional servers & Provision and evaluation  of power metering functionality for virtualized data centers. \\ \hline
     \cite{LeBeane2015} & Creating a measurement model that is sensitive to a number of components & Other power measurement approaches are not detailed enough to be useful for chip design & Proposal and validation of measurement framework\\ \hline
     \cite{Henderson2020} & Furthering the ability to measure and report energy and carbon intensity of machine learning models & Available methods for collecting energy and carbon data are not widely adopted & Providing an easy to implement measurement framework \\ \hline
     \cite{Jagroep16} & Tracking a software products energy consumption over multiple releases & Difficulties in relating software changes to changes in power consumption & New regression model for power consumption estimation and comparison with existing model. Contextualizing measurements with software changes. \\ \hline
     \cite{Kreten2022} & Extensive dissertation on the investigation of energy efficient use of container technology & Using containers to improve energy efficiency of data centers & Presentation of measurement method, environment and guidelines as well as supporting tools for energy efficient use. Achieved improvements in 7 out of 10 use cases. \\ \hline
     \cite{Verdecchia2018} & Identifying code sections causing high energy consumption & Difficulties in determining which parts of code induce how much energy consumption, particularly which parts induce disproportionate consumption & Proposed method to determine induced energy consumption by individual functions, code branches or lines using experimental validation \\ \hline
     \cite{GSL2021} & Project to further the establishment of green software engineering as a discipline. Locating code section causing excessive energy consumption & Lack of tool support & Development of several tools and frameworks for identifying energy hot-spots and optimizing source code for energy efficiency\\ \hline

        \hline
   
\end{longtable}
\end{landscape}

In addition to the content found in Tab.~\ref{tab:rq1_label}, it is important to the authors that purely technical possibilities, which are not (yet) published in papers but are known to the community, are also presented. For example, a natural step to optimize the energy consumption of software is continuously measuring power consumption during development. Through continuous measurements, changes in energy consumption between builds and deployments may be attributed to concrete changes or code sections. This information may then inform optimization efforts e.g. by changing libraries, call hierarchies, structures, frameworks, algorithms, or even SDK updates, etc. Making this transparent to the developers, but also other relevant stakeholders like users, is the first step to a sustainable software product.

In addition to the blue angel for software products and the criteria developed in~\cite{kern18}, as well as the the proposed measurement setups using external power meters, internal loggers (like the aforementioned RAPL and nvidia-smi or eBPF\footnote{\url{https://ebpf.io/} [2023-04-28]}) in combination with resource loggers can go a long way in assessing the impacts of a software product under development. To be able to access the data provided by the loggers, it is important which operating system is used. it is also necessary to access the data provided by the hardware drivers. The following list contains current middleware for simplified access to the information provided are on the one hand for RAPL
\begin{itemize}
    \item Green Metrics Tool\footnote{https://github.com/green-coding-berlin/} by Green Coding Berlin GmbH\footnote{https://www.green-coding.berlin/}
    \item Hubblo\footnote{https://hubblo.org/} RAPL exporter for Windows\footnote{https://github.com/hubblo-org/windows-rapl-driver} using Prometheus\footnote{https://prometheus.io/}
    \item Hubblo energy consumption metrology agent Scaphandre\footnote{https://github.com/hubblo-org/scaphandre} using Prometheus metics
\end{itemize}
and on the other hand for eBPF
\begin{itemize}
    \item Kepler (Kubernetes-based Efficient Power Level Exporter)\footnote{https://github.com/sustainable-computing-io/kepler} by Sustainable Computing\footnote{https://sustainable-computing.io/}. This tool exports Prometheus metrics.
\end{itemize}

\subsection{RQ2: What concepts of environmentally friendly software engineering processes are already in place in the Internet industry?}
Few case studies on software sustainability practices in the industry are available. Calero et al.~\cite{calero21a} performed an in-depth analysis of software sustainability on a set of personal health record software products. They propose recommendations and measured their impact on software energy consumption.  Heldal et al.~\cite{heldal2023} identify a need for sustainability competency in IT professionals in the industry. Lammert et al.~\cite{lammert2022} found that software engineers are usually not equipped with the training and tools to fulfill this need. In addition to scientific papers, industry and developer communities have also produced methods and tools for sustainable software engineering, some of which are not (yet) published. These often build on the software measurement approaches mentioned in RQ1.

In addition to the blue angel for software products and the criteria developed in~\cite{kern18} and the proposed measurement setups using external power meters like~\cite{junger22}, internal loggers (like the aforementioned RAPL and nvidia-smi or eBPF\footnote{\url{https://ebpf.io/}~[2023-04-28]}) in combination with resource loggers can be useful to assess the impacts of a software product under development. Here, it is necessary to access the data provided by the hardware drivers. Exemplary tools are the Green Metrics Tool\footnote{\url{https://github.com/green-coding-berlin/}~[2023-05-09]}, Code Carbon\footnote{\url{https://github.com/mlco2/codecarbon}~[2023-05-10]} Hubblo RAPL exporter\footnote{\url{https://github.com/hubblo-org/windows-rapl-driver}~[2023-05-09]} and Scaphandre\footnote{\url{https://github.com/hubblo-org/scaphandre}~[2023-05-09]}.

Many available guidelines and best-practice recommendations are in the form of blog posts or lists from developer communities, consulting firms and businesses\footnote{\eg \url{https://www.ibm.com/cloud/blog/green-coding}, \url{https://www.suso.academy/en/2023/03/13/green-coding-the-5-most-important-basics-for-sustainable-software-development-with-code-examples/}, or \url{https://geekflare.com/green-coding/} [2023-06-27]}. Regardless of the correctness of the information contained in this content, they often do not provide sources for their recommendations.

\begin{landscape}
\setlength\LTcapwidth{\linewidth}
\begin{longtable}[h]{|m {4em}|m {12em} |m {15em} |m {16em}|}
    \caption{Analyzed Papers regarding to RQ2: What concepts of environmentally friendly software engineering processes are already in place in the Internet industry?}
    \label{tab:rq2_table}\\
    \hline
    \textbf{Source} & \textbf{Research Objective} & \textbf{Problem or Gap Addressed} & \textbf{Findings/Conclusion}\\
    \endfirsthead
    \hline 
    \textbf{Source} & \textbf{Research Objective} & \textbf{Problem or Gap Addressed} & \textbf{Findings/Conclusion}\\
    \endhead
    \hline 
\cite{khan18} &
Experiences in Using RAPL for Power Measurements
 &  
Power measurement using the RAPL  & 
Measurement with RAPL is possible\\
 \hline
        
\cite{acar17} &
Software development methodology in a Green IT Environment
 &  
A doctoral thesis about the software development methodology in a Green IT environment
 & 
The main goal of this paper is to provide a methodology allowing the creation of sustainable and green software products to reduce greenhouse gas emissions.
\\

\hline
        
 \cite{muthu19} &
Green and sustainability in software development life-cycle process - Greensoft Model & 
How sustainable software development can be implemented
 & 
The GREENSOFT model has explained three order effects: the first-order effect focuses on development to end, the second-order concentrates on sustainability area, and the last focuses on recommendations and tools for IT users in creating and maintaining a product for sustainable development.
\\
 \hline
 
\cite{kramer18} &
Best practices in systems development lifecycle: an analysis based on the waterfall model &  
There is a process known as Systems Development Lifecycle (SDLC), which can make the process both simpler and less stressful. & 
Some best practices of a needs analysis are accomplished with the following: meeting with Stakeholders as a group or one-on-one to find out their requirements, discussing their current challenges, and retrieving any documents containing current processes.
\\
 \hline
 
 \cite{ahmad22} &
Green Software Process Factors: A Qualitative Study &  
For instance, working on many other features in the same period and delaying testing is a fundamental waste-related problem. Additionally, rework is frequently wasted during development, correcting work to be delivered but not done appropriately.
 &
This study has revealed research on the software process, green software process, software sustainability, and software waste.
\\
 \hline 
 
 \cite{zhang10} &
Cloud computing: state-of-the-art and research challenges &  
the development of cloud computing technology is in its infancy, with many issues still to be addressed.
 &
In this paper, the author has surveyed state-of-the-art cloud computing, covering its essential concepts, architectural designs, prominent characteristics, key technologies, and research directions. As the development of cloud computing technology is still at an early stage, we hope our work will provide a better understanding of the design challenges of cloud computing and pave the way for further research in this area.
\\
 \hline 
 
 \cite{kern15} &
Labeling sustainable software products and websites: ideas, approaches, and challenges &  
The demand and problems of a sustainability label for software products and its engineering need to be discussed in general.
 &
The aspects that are relevant while creating a label for sustainable software products
\\
 \hline 
 \cite{JOUMAA12} &
Green IT: Case Studies &  
Recently, studies have been conducted to identify the main pollutant components in IT systems.
 &
In this paper, the author focused on the consumption of energy by computers. We presented two case studies that help in greener computing. In the first case study, we proved that using thin clients instead of PC results in considerable savings in terms of energy consumption. In the second case, we proposed a proxy-based Energy Efficient BitTorrent architecture that can enhance energy savings without affecting the QoS.
\\
 \hline 
 
 \cite{dick10} &
Green Web Engineering: A set of principles to support the development and operation of “Green” websites and their utilization during a website’s life-cycle &  
The power consumption of ICT and the Internet is still increasing. To date, it is still being determined whether the energy savings through ICT overbalance the energy consumption by ICT. In either case, it is the suggestion to enforce the energy efficiency of the Web.
 &
The power consumption of ICT and the Internet is still increasing. However, whether or not energy savings through ICT overbalance its energy consumption needs to be clarified. In either case, it is rational to enhance the energy efficiency of the Web.
\\
 \hline 
\cite{thilakarathne22} &
Green Internet of Things: The Next Generation Energy Efficient Internet of Things &  
involving IoT devices drastically increases the data sensor, and actor data shared throughout the internet. The Green IoT envisages reducing the energy consumption of IoT devices and keeping the environment safe and clean. overview of Green IoT (GIoT) &
GIoT could offer many advantages, such as environmental sustainability and protection, end-user satisfaction in different IoT domains, and minimizing the harmful effects on the environment and human health \\
 \hline 
 \cite{Psannis14} &
Convergence of the Internet of Things and mobile Cloud computing &  
The paper explores how mobile cloud computing (MCC) can improve the Internet of Things (IoT) limitations. IoT is a real-world interaction with everything connected via intelligent network infrastructure, while MCC offers an infrastructure for data storage and processing outside of the mobile device. The paper presents the advantages of MCC over IoT and highlights the benefits of integrating the two &
 critical assessment of various MCC advances regarding specific IoT green applications constitutes future research direction. Finally, a future goal is the cross-discipline technology integration to validate the convergence of green IoT and MCC framework.\\
 \hline 
\end{longtable}
\end{landscape}

\subsection{RQ3: How can Green Coding concepts be implemented into the current software development processes and the curricula of existing study programs?}

\subsubsection{RQ3-1: How can Green Coding concepts be implemented into the current software development processes?}

Firstly, motivators and reasons for companies are displayed to answer the question of how Green Coding concepts are being integrated into the current software development process. The first reason is obviously to reduce the energy consumption. Considering the world's energy production, which consists of non-renewable and renewable energy sources, one can assume a direct link to carbon dioxide emissions. The composition of a country's electricity mix varies widely and directly influences the emissions per consumed unit of energy. This means that a reduction in energy consumption can be directly linked to decreased $\text{CO}_2$ emissions. Based on the assumption that the required electricity must also be purchased, it can be assumed that a reduction in electricity consumption will also reduce electricity costs. This effect will be most noticeable for companies operating data centers.

In addition to the costs that can be saved through energy consumption, it must also be assumed that more efficient algorithms will reduce operating resource consumption. This leads to a reduction in the hardware required due to more efficient software and, in turn, to lower hardware overhead in the case of rescaling or, in the case of no rescaling, to a longer service life of the hardware. In both cases, this can be directly linked to cost savings. 

In addition to the cost savings from more efficient hardware and energy use, the reduction of direct and indirect environmental impacts is in the foreground in the Green Coding context. Only by tracking and, if necessary, improving can environmental guidelines and environmental regulations be complied with. Ultimately, environmental seals, labels, and awards can be sought that can provide a competitive advantage. For example, an eco-label is a clear advantage for public procurement in Germany, as it is can be part of the specification for a product \cite{Schneider2023}. However, even without regulations, it can be assumed that a green impact will benefit from transparency. 

To complete the reasons for implementation in the current software development cycle, the authors have considered where the knowledge can be implemented in companies. So the question must be asked how the knowledge is transferred to the corresponding places. In this context, training and education are fundamental anchor points for transfer and implementation. 

On the one hand, the conceptual and technical basics of Green Coding or green it, in general, should be conveyed. Training courses and workshops must bring this knowledge into the companies. These training courses can be given or initiated externally by consultants and governmental actors. However, they can also be quickly and widely disseminated through e-learning platforms.

To take advantage of the internal scaling effect, it would be advantageous if a person or team responsible for this area first attended these training courses and then passed them on to the relevant departments in the form of internal training courses. This would be a possible way to implement a cooperate standard that provides incentives and obligations for tracking and implementing energy and resource consumption. In this way, governmental and non-governmental actors or even consulting companies can also provide templates for internal training programs. 

Finally, after creating a particular basis, knowledge can be strengthened through conferences and meetups. An enormous exchange of knowledge occurs, especially on the fringes of conferences, which discuss and lecture on the latest development trends in the industry. To implement the path most effectively, investing in tools and frameworks for implementing code or adapting the software development process is indispensable. Only by tracking and informing about the current usage and possible alternatives can the developing software engineers be given a fast and quick impression and a suggestion of better solution possibilities.

The following Table \ref{tab:rq3-1_table} displays the literature regarding implementation in the software development process between 2010 and 2022.

\begin{landscape}
\setlength\LTcapwidth{\linewidth}
\begin{longtable}[h]{|m {4em}|m {12em} |m {15em} |m {16em}|}
    \caption{Analyzed Papers regarding RQ3-1: How can “Green Coding” concepts be implemented into the current software development processes?}
    \label{tab:rq3-1_table}\\
    \hline
    \textbf{Source} & \textbf{Research Objective} & \textbf{Problem or Gap Addressed} & \textbf{Findings/Conclusion}\\
    \endfirsthead
    \hline 
    \textbf{Source} & \textbf{Research Objective} & \textbf{Problem or Gap Addressed} & \textbf{Findings/Conclusion}\\
    \endhead
    \hline 
\cite{dick10-2} &
Sustainability in software and software engineering & 
Definition of Sustainable Software and Sustainable Software Engineering & 
Generalization of a sustainable software engineering process. The authors add sustainable criteria to a generalized approach. Sustainability reviews \& previews, process assessments, sustainability journals, and sustainability retrospectives are explained further. The authors map these ideas in sequential and iterative development processes.\\
 \hline

\cite{ibrahim21} &
Green software engineering with agile methods & 
Include Green IT aspects into software engineering processes with agile methods to produce “greener” software from scratch & 
Integration of sustainability aspects in Scrum\\
 \hline

 \cite{kocak13} &
Green software development and design for environmental sustainability &  
Focuses on investigating the trade-off between software quality requirements and environmental sustainability & 
Summarizing the knowledge and the experience collected within the exploratory study. \\
 \hline
 
 \cite{naumann21}&
Development and components of the eco-label Blue Angel for software &  
This article presents the “Blue Angel”, a label evaluating and classifying software's resource and energy efficiency. In particular, the process by which the label was developed is presented.& 
The paper describes the components of the Blue Angel and the focus of the first iteration of the Blue Angel for energy and resource-efficient software products.\\
 \hline 
 
 \cite{swacha22}&
Models of Sustainable Software: A Scoping Review &  
n this paper, we present the results of a scoping review of the literature on sustainable software models based on 41 works extracted from an initial set of 178 query results from four bibliographic data providers. &
The key contribution of the reported research is mapping existing literature on green and sustainable software models using five categories (model scope, purpose, covered sustainability aspects, verification or validation, and the economical category of the research country).\\
 \hline 
 
 \cite{lavanya20} &
GREEN SCRUM MODEL: Implementation of Scrum in Green and Sustainable Software Engineering &  
Agile methodology is one of the software practices that entire software companies revolve around. Sustainable development is the idea that human societies must live and meet their needs without compromising the ability of future generations to meet their own needs. &
The main objective of the proposed GREEN SCRUM Model is to structure concepts, strategies, activities, and processes of the Scrum model and especially of green and sustainable software and its engineering.\\
 \hline 
 
 \cite{yarlagadda19} &
The DevOps Paradigm with Cloud Data - Analytics for Green Business Applications &  
In this paper, the aim is to understand how the DevOps framework works in green business applications using cloud-based software solutions &
The findings described in this paper have important applications in IT management and information analytics, but there are yet unresolved issues\\
 \hline 
 
 \cite{jeya20} &
Towards Green Software Testing in Agile and DevOps Using Cloud Virtualization for Environmental Protection &  
The objective of this chapter is twofold: firstly, to provide a green software testing framework using cloud-based virtualization, and secondly, to apply cloud-based testing in Agile and DevOps-based software development environments. &
To achieve greener effects in the software development process in DevOps and Agile, the cloud-based solution helps in monitoring the changes incorporated in the software using a version control system provided in the cloud of the organization\\
 \hline 
 
 \cite{betz14} &
Sustainable Software System Engineering &  
until now, sustainability is not considered by software system engineering. Hence, to support the transition to sustainability, one must have sustainability (aspects) integrated into the software systems and the underlying business processes. &
So far only single solutions for example for carbon emissions, business processes, or green software engineering have been developed. None of the approaches has a holistic approach, and none considers the software systems and underlying business processes.\\
 \hline 

 \cite{kern13} &
Green Software and Green Software Engineering – Definitions, Measurements, and Quality Aspects &  
The paper describes a reference model for green and sustainable software, as well as its engineering, and also gives some definitions. &
In this paper,  a model is presented to classify green software and its engineering and shows some aspects of how the energy efficiency of software can be measured.\\
 \hline 
 \cite{Sus_Naumann2011} & 
 Green Software Engineering with Agile Methods &  
The rising energy consumption of ICT is worrying, and although hardware solutions for Green IT exist, the impact of software on energy consumption remains unexplored. The paper presents a model integrating Green IT into software engineering processes using agile methods to create eco-friendly software. & 
The paper presents a model integrating sustainability issues into software development, incorporating Green and Sustainable Software Engineering due to a sustainability report into Scrum. The model supports engineers in creating eco-friendly software, although its effectiveness cannot be verified without measuring and monitoring software performance regarding these issues.\\
 \hline
 
  \cite{Greens_Dick2013}  &
 The GREENSOFT Model: A reference model for green and sustainable software
and its engineering &  
Academic research has studied the relationship between sustainable development (SD) and information and communication technology (ICT). The focus has been on the impact of ICT on environmental sustainability and the balance between ICT's energy and resource savings versus its consumption. However, there is no consensus or definition on whether ICT's energy and resource savings will outweigh its consumption.
 & 
 The paper proposes definitions of "Green and Sustainable Software" and "Green and Sustainable Software Engineering". It outlines a conceptual reference model, the GREENSOFT Model, which includes sustainability metrics and criteria for software and software engineering extensions for sustainable software design and development.\\
 \hline
  \cite{bal23}  &
Tactics for Software Energy Efficiency: A Review
 &  
The study aims to understand the state of energy-efficient tactics for software systems, using a systematic literature review and analyzing 142 primary studies on 163 tactics. The research interest peaked in 2015 but declined afterward, with a focus on source code static optimizations and application-level dynamic monitoring. Limited industry involvement hinders practical application, emphasizing the need for collaboration between academic researchers and industrial practitioners to make a real impact on energy efficiency.
 & 
The paper aims to analyze software energy efficiency tactics through a literature review, identifying 142 primary studies and 163 tactics. Despite growing energy consumption concerns, the research area faces declining interest and limited industrial involvement. The call is for collaboration between academics and industry practitioners to address software sustainability effectively.\\
 \hline
 
    \end{longtable}
\end{landscape}

\subsubsection{RQ3-2: How can Green Coding concepts be implemented into the curricula of existing study programs?} 

To answer the question of how Green Coding or Green Software Engineering can be integrated into universities or institutes of higher education in the future and to be able to make good to excellent recommendations, the indispensable first step is to research existing courses and programs. However, the search itself is more challenging than it may initially seem. Most universities have printed the individual specializations of the study subjects in so-called module handbooks or make the course information with their disciplines available only after an account has been created in a campus management system.

This can influence the search, which was mainly done via online media. In addition to sources from cooperating universities, job platforms such as LinkedIn\footnote{\url{https://www.linkedin.com/}} and institutions known in the Green Coding community had to be used to achieve the most meaningful result. Currently, courses specifically designed for Green Software Engineering or Green Coding can be found mainly in Europe. The following universities and universities of applied sciences in Europe teach related courses.

\begin{figure}
    \centering
    \includegraphics[width=10cm, height=10cm]{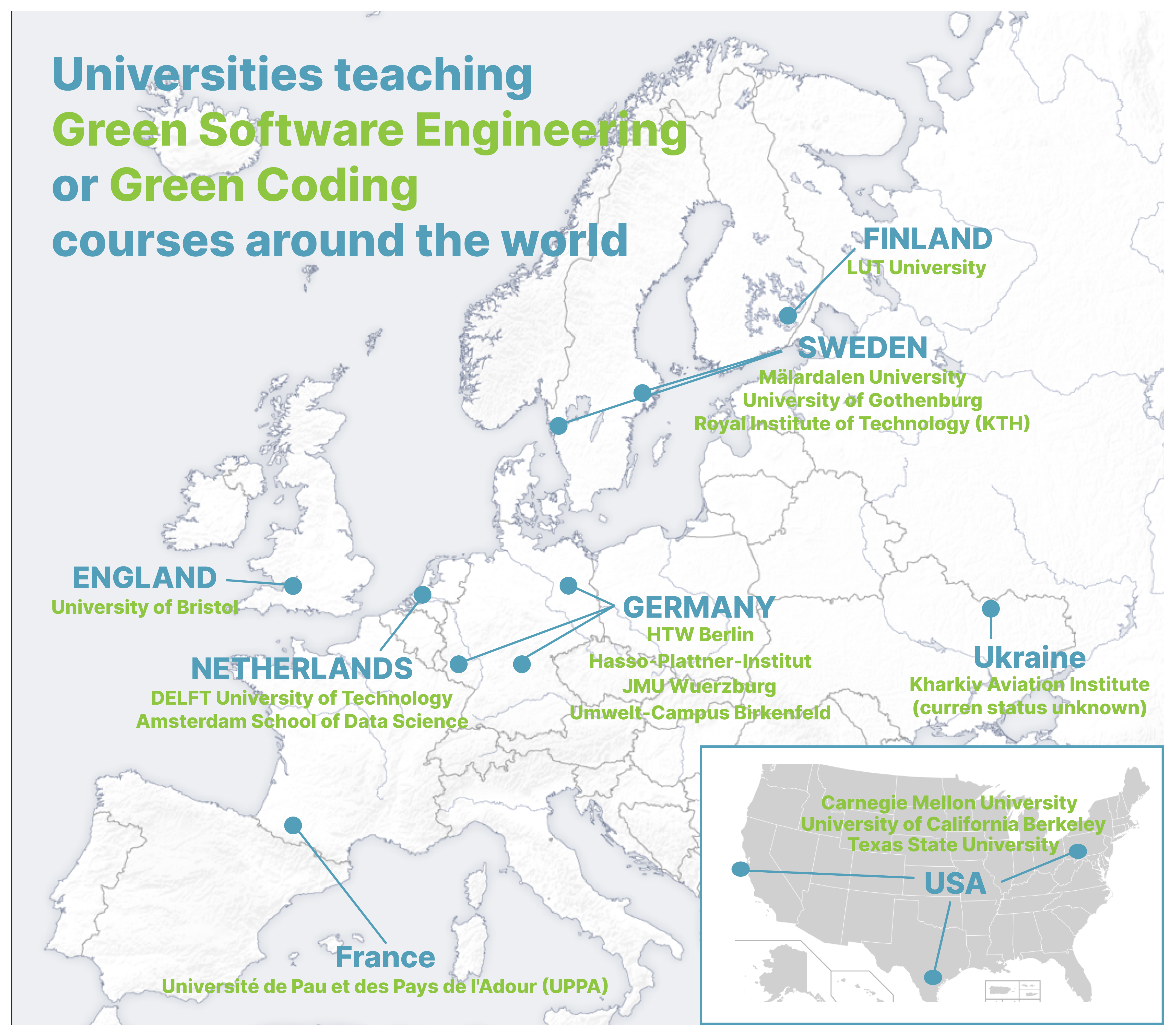}
    \caption{Universities teaching Green Software Engineering or Green Coding courses around the world 
}
    \label{fig:teaching_universities}
\end{figure}

\begin{itemize}
\item England - University of Bristol
\item Finland - LUT University
\item France - Université de Pau et des Pays de l'Adour (UPPA)
\item Germany - HTW Berlin,  University of Applied Sciences
\item Germany - Hasso-Plattner-Institut (HPI)
\item Germany - JMU Wuerzburg
\item Germany - Umwelt-Campus Birkenfeld (UCB)
\item Netherlands - DELFT University of Technology 
\item Netherlands - Amsterdam School of Data Science
\item Sweden - Mälardalen University
\item Sweden - Royal Institute of Technology (KTH)
\item Sweden - University of Gothenburg
\item Ukraine - Kharkiv Aviation Institute (current status unknown)

\end{itemize}
The following universities were identified within the United States of America.
\begin{itemize}
\item Carnegie Mellon University
\item the University of California Berkeley
\item Texas State University
\end{itemize}
The universities mentioned in figure \ref{fig:teaching_universities} were graphically processed to get a better spatial picture of the distribution.
To be able to give recommendations for the implementation of 'Green Coding' and 'green software engineering', in addition to the search terms listed in the table \ref{tab:Seach_Strings}, very explicitly necessary search terms for the creation of curricula are added to the analyzed literature.  More detailed analyses of the selected sources can be found in the following table \ref{tab:rq3-2_table}.

\begin{landscape}
\setlength\LTcapwidth{\linewidth}
\begin{longtable}[h]{|m {4em}|m {12em} |m {15em} |m {16em}|}
    \caption{Analyzed Papers regarding RQ3-2: How can “Green Coding” concepts be implemented into the curricula of existing study programs?}
    \label{tab:rq3-2_table}\\
    \hline
    \textbf{Source} & \textbf{Research Objective} & \textbf{Problem or Gap Addressed} & \textbf{Findings/Conclusion}\\
    \endfirsthead
    \hline 
    \textbf{Source} & \textbf{Research Objective} & \textbf{Problem or Gap Addressed} & \textbf{Findings/Conclusion}\\
    \endhead
    \hline
 \cite{batlegang15} &
How can environmental issues be incorporated into the existing IT curriculum at Botho University (Botswana)& 
Lack of Green Computing courses & 
Levels of Green Computing awareness at university are low, and there are minimum green initiatives \\
 \hline
        
 \cite{bovill19} &
Co-creation of the curriculum in higher education
 & 
Not enough focus on the inclusion of students in curriculum development 
 & 
More collaborative form of curriculum benefits educators as well as students
\\
 \hline
 
 \cite{drake18} &
Development of an integrated curriculum
 & 
Teaching expertise should be the central focus of the curriculum.
 & 
The curriculum should provide the big picture and present competencies as a backdrop for unit plans.
\\
 \hline

 \cite{sharon06}  &
Definition of the Curriculum in a higher education context
 &  
Ambiguous term in higher education with variation in perceptions
 & 
Curriculum determined by technical, practical, and participatory interest.
\\
 \hline
 
 \cite{ramirez18} &
Development of a new curriculum for engineering education for Industry 4.0
 &  
Engineering education must be adjusted regarding Industry 4.0
 &
Interdisciplinary knowledge/competencies are more critical than ever.
\\
 \hline 
 
 \cite{knight01} &
Process model curriculum creation
 &  
Curriculum coherence is not widespread.
 &
Less focus on outcome thinking (faithful translation of plans into practices). More focus on program learning processes.
\\
 \hline 
 
 \cite{lozano14} &
Creation of a curriculum in engineering for sustainable development
 &  
The growing interest in sustainability and educated and empowered students
 &
Integrated curriculum - sustainability intertwined as a concept in all courses
\\
 \hline 
 \cite{lubicz18} &
Benefits of co-creation of curriculum
 &  
Ever-changing complex world - current curriculum power dynamics won‘t allow the necessary personal growth. 
 &
Co-creation of the curriculum can improve self-authorship and cognitive and interpersonal abilities.
\\
 \hline 
 
 \cite{malik11} &
Twelve tips for Developing an integrated curriculum
 &  
Knowing about different levels of integration makes it easier to develop a new curriculum.
 &
The difficulty of change in established curricula - discussions, and workshops as an initial step
\\
 \hline 
 
 \cite{somekh09} &
Action research approach for curriculum development
 &  
Traditional teaching techniques fail to encourage „deep“ learning of subject content.
 &
Students benefit from an action research cycle with a practical reflection upon one's experiences publicly. 
\\
 \hline 
 \cite{ryan13} &
Flexible pedagogical ideas in higher education
 &  
Need for flexibility in higher education to adapt to changing learning landscape.
 &
Delivery of 6 new pedagogical ideas concerning the way subjects are taught as well as institutional implications
\\
 \hline    
  \cite{Torre2017} &
On the Presence of Green and Sustainable Software Engineering in Higher Education Curricula
 &  
A Survey is created to determine the current state of green and sustainable software engineering in higher education. The survey findings are discussed and analyzed.
 &
Besides the integration in the main streams of green or sustainable software engineering, the authors want to set up in-depth interviews with higher education stakeholders to identify more insight about the educator's view with the target to develop a concrete integration idea about topics, knowledge guidelines, and course content.
\\
 \hline    
  \cite{Saraiva2021} &
Bringing Green Software to Computer Science Curriculum: Perspectives from Researchers and Educators
 &  
Research shows software engineers lack the knowledge, expertise, and tools to develop eco-friendly software. Recent studies indicate green software should be taught in modern computer science curricula. Surveys are used to find a solution.
 &
Green software in computer science curricula can benefit the planet, society, and students. However, the current curriculum lacks teaching materials and innovative education in green computing. The paper presents survey results on green software education, key pedagogical issues, and exemplary solutions. Bringing green software to the computer science core curriculum requires a long-term community effort. Stakeholders from industry and academia are invited to join in promoting green software design.
\\
 \hline    
   \cite{Mishra2021} &
Sustainable Software Engineering: Curriculum Development Based on ACM/IEEE Guidelines
 &  
How sustainability can be included in various courses of the SE curriculum by considering ACM/IEEE curriculum guidelines for the SE curriculum
 &
The Mishras identify Core Sofware Engineering courses already implementing or should implement the Green Coding techniques in the future, e.g.,  Requirement Engineering, Architecture and Design, HMI Design, and Cloud Computing, and ranks them by kind and art. It displays key competencies that should be teached to under- and graduate students.
\\
 \hline 
   \cite{eickstaedt2023} &
Computer Science for Future Sustainability and Climate Protection in the Computer Science Courses of the HAW Hamburg
 &  
The aim of the initiative is a paradigm shift in the discipline of computer science, thus establishing sustainability goals as a primary leitmotif for teaching and research.
 &
Teaching sustainability is a multiplier for spreading sustainability by teaching university students. The change will impact research and a transfer to business, academics, and civil society.
\\
 \hline 
  \cite{Turkin2019} &
The paper presents an academic case where the authors describe the elaborated design of the course “Software Engineering Sustainability” to introduce the sustainable concept into educational programs for software engineers in the master's degree program 'Software Engineering' of the Kharkiv aviation institute.
 &  
The presented course design considers the requirements of employers to maintain the successful experience of national higher education and professional industry standards and conforms to the European Union e-Competence Framework. 
 &
The authors present the design of the course with a description of the course content and visible positioning of Green IT. The course has been delivered by the first author at National Aerospace University “Kharkiv Aviation Institute”.
\\
 \hline 
  \cite{cai2010} &
This paper advocates for integrating sustainability into undergraduate computing education, presenting three sustainability integration strategies, a green computing course, and learning modules to prepare graduates with competencies to create a sustainable future.
 &  
Computing graduates are under-prepared in topics like Waste Management, eco-labeling, energy management, carbon management, reuse, environmental reporting, and open-source software.
 &
 Yun creates a hypothetical agenda for integrating computing education for sustainability.
\\
 \hline 

  \cite{Kor2019} &
Education in Green ICT and Control of Smart Systems: A First-Hand Experience from the International PERCCOM
Masters Programme 
 &  
PERCCOM is the first international Green ICT Master's program to train new IT engineers in sustainable digital application design and implementation. The paper assesses the skills and employability of PERCCOM graduates after five years of the program and offers recommendations for environment-related education curricula.
 &
PERCCOM is the first Masters's program on Green education in digital sectors globally, with positive feedback. However, survey results reveal the general need for Green IT in companies where PERCCOM Alumni work, perhaps due to the acute shortage of qualified automatic control systems and ICT engineers in Europe. Though data centers recognize energy efficiency as a significant challenge, automated control systems consume substantial resources, and future controllers should be evaluated based on sustainability metrics and traditional ones. This consideration should be included in automatic control curricula.
\\
\hline 

  \cite{Angelaki23} &
Towards more sustainable higher education institutions: Implementing the sustainable development goals and embedding sustainability into the information and computer technology curricula
 &  
This paper discusses integrating sustainability issues and sustainable development goals (SDGs) into higher education, specifically in the Information and Communication Technology (ICT) field. It addresses the challenges of introducing education for sustainability in higher education institutions, proposes principles for incorporation, and advocates for sustainability integration in ICT curricula. The study at a Greek university shows that educational intervention positively influences students' intentions to engage in sustainability, enhances their understanding of sustainability, and highlights the need for more inclusive awareness activities to support sustainability engagement.
 &
 
The paper deals with the general integration of sustainable development (SD) into higher education. A holistic approach is chosen to include sustainability in curricula, as this is the only way to ensure sustainable development. positive results were found after an experimental study at a university in Greece. These results are related to increased student awareness and, thus, the importance of integrating sustainability issues into ICT curricula. The paper argues for sustainable education practices in higher education institutions to produce a new generation of ICT professionals who can effectively deal with sustainability issues.

\\
 \hline   
    \end{longtable}
\end{landscape}

In curricula to be prepared, on the one hand, the big picture of a topic and, on the other hand, the main competencies should be taught \cite{drake18}. To convey this knowledge, technical, practical, and participatory interests should be addressed \cite{sharon06}. 
How a sustainable software engineering curriculum based on ACM/IEEE could be created is described by \cite{Mishra2021}. Since it is difficult to design entire curricula such as PERCCOM \cite{Kor2019}, the first step should be to create workshops, discussion groups \cite{malik11} or university initiatives like in the HAW Hamburg \cite{eickstaedt2023}. This can be followed by a module design like 'Software Engineering Sustainability' of the Kharkiv aviation institute (\cite{Turkin2019} (current status cause of war unknown). To identify the right target group for these modules, it can be assumed that if it is implemented as a module, it should be attended by various similar courses, because interdisciplinary knowledge is more important than ever in current times \cite{ramirez18}.
\subsection{Creation of a Green Coding pilot course and integration into the existing curriculum at the HTW Berlin}
In the context of the third research question, which was led by HTW Berlin, the conception of a module in the Master's program 'Industrial Environmental Informatics' took place as a pilot project. A test integration at HTW Berlin implemented the concept. The teaching subject "Current Development Trends in Environmental Informatics" now introduces students to the basic concepts and techniques of environmentally conscious and energy-efficient software development. The course focuses on teaching students how to develop software that is not only energy efficient but also environmentally friendly. It teaches how software developers can optimize the energy efficiency of software applications by, for example, writing energy-efficient algorithms or optimizing the performance of servers and data centers. For this purpose, the basic concepts of 'Green IT' are taught, and the subfield of Green Coding is addressed. 
The course is divided into three modular parts: internal lectures, external lectures, and a practical part to manifest the theoretical knowledge. The content of the internal lectures was significantly influenced by the current state of research regarding 'Green IT' at HTW Berlin, in which current and past master theses, papers, and publications were presented. External researchers gave lectures, partly on-site and online. For example, experts from the 'Öko-Institut~e.V.' gave lectures on the life cycle costs of ICT, the digital footprint of ICT, and the Blue Angel for software products. The LMU Munich presented the basics of Green IT in data centers and 'Green Coding Berlin GmbH' showed the current state of the art in the software industry. To learn more about the topic of implementation, the 'Greensoft Model' was presented by the participating researchers of the UCB, with a focus on 'Green Software' and 'Green Software Engineering'.
To be able to give the most comprehensive grade possible, which is reasonably differentiated and represents the participating teachers developing all the skills taught, a multi-level grading system, as a prerequisite for admission to the final exams, the students had to assess the effects of ICT hardware. In this context, the 'Oeko-Institut~e.V.' introduced the ''digital carbon footprint'' online calculator with a life cycle cost module. For this purpose, the necessary calculation was to be made and designed from the data supplied for a digital product. The purpose was to understand the direct and indirect effects of hardware on the environment, in the form of materials used or energy consumed, and their impact and to apply the knowledge directly. By publishing the results on the https://www.digitalcarbonfootprint.eu/ website, which was used as a multiplier for this task, the results can immediately be passed on to a broad audience.

Further grading was divided into a theoretical and a practical part. The academic content was examined through oral examinations focusing on 'Green IT', 'Green Computing' and 'Green Coding'. A measurement laboratory modeled on the Blue Angel for software products was designed to test the functional performance. With the help of the UCB, which like the HTW had a scientist on-site in each case, the students could pursue their research question in this collaborative project.

Based on the previous master thesis about measuring the power consumption of LCA Software \cite{junger22} and the preliminary work of the Blue Angel for software products, a software measurement was carried out by the students, who also implemented the experiments. Topics included comparisons of 2 products/frameworks/SDKs to assess the energy and resource comparison of current
\begin{itemize}
\item applications focusing on statistical analyses and evaluations using the example of R / Matlab,
\item programming languages using the example of functional and object-oriented programming using C and Python,
\item applications with the emphasis on music processing, using Spotify and VLC Media Player as examples,
\item applications with the main focus on image processing, using GIMP and Photoshop as examples, and
\item applications focusing on single-threading and multi-threading using the Go programming languange.
\end{itemize}

The compiled results were presented to the entire course as lessons learned and written down as a 5-page report. A future publication of the best paper at a scientific conference is pending.

Besides the sensitization and the transfer of the Green Coding basics and the general knowledge about improvement possibilities, the studies could be evaluated continuously. In self-assessments at the beginning and at the end of the course, content-related aspects were controlled and a continuous improvement in all queried fields could be determined. An elaboration of this pilot project and the evaluation results will be presented at the 53rd annual conference of the "German Informatics Society" in the "environmental informatics for the design of a sustainable future" workshop.

\clearpage

\section{Conclusion}
In this report, we collected results from an exploratory literature review to classify the state of the art of 'Green Coding' in software engineering and the education classified as support in software engineering with studies, research, discussions, etc., to find out the limits, ways and means that exist. This should enable stakeholders to realize the potential of Green Coding, whether well-researched or untapped and provide them with a solid introduction to the field and possible approaches for their activities. Whereas \cite{ease14} concluded that the area of \textit{Green IT} is still in its infancy, as evidenced by a lack of case studies and evaluations, we conclude that research in the field is steadily increasing and has made significant progress. Several measurement and evaluation systems for assessing software's energy- and resource consumption and -efficiency have since been developed and published. However, the point about a low adoption rate of Green IT methods still stands, especially in the Internet industry, as few case studies have been identified. Implementing aspects of sustainability, such as Green Coding techniques, into the training of future IT professionals is a crucial step to bringing sustainability into view for software producers while providing them with a workforce capable of implementing sustainability into their development process.

Future work could be done by extracting practices contained in non-academic sources and finding academic papers supporting their impact on sustainability or attempting to prove some of the recommendations experimentally by applying them to a software product and tracking the impact of each recommendation. Surveys and expert interviews should be conducted to get a better overview of green coding in institutes of higher education and industry, as curricula may not be up-to-date and corporations do not always publish the results of their projects.

A logical consequence of this project is that this knowledge is listed and communicated in a way that is as easy to understand as possible. Besides already existing courses, like the course from the green software foundation \footnote{\url{https://learn.greensoftware.foundation/}~[2023-05-09]}, it is essential that these concepts are scientifically validated and distributed. Since the expertise in imparting knowledge lies with the institutes of higher education, the design for the course concept to impart this knowledge and practice needs to be expanded.

\section*{Acknowledgements}
This work was funded by the Internet Society Foundation\footnote{\url{https://www.isocfoundation.org/}~[2023-11-30]} project ''Potentials of Green Coding``. We would also like to thank our partner, the German Informatics Society (Gesellschaft für Informatik e.~V.), especially Elisabeth Schauermann and Carolin Henze\footnote{\url{https://gi.de/en/aktuelles/projekte/en-green-coding}~[2023-11-30]}, the assisting master students, and student assistants.

\bibliography{Potentials_of_Green_Coding} 

\end{document}